\documentclass[fleqn,usenatbib]{mnras}

\usepackage{newtxtext,newtxmath}

\usepackage[T1]{fontenc}

\DeclareRobustCommand{\VAN}[3]{#2}
\let\VANthebibliography\thebibliography
\def\thebibliography{\DeclareRobustCommand{\VAN}[3]{##3}\VANthebibliography}

\usepackage{graphicx}	
\usepackage{amsmath}	
\usepackage{subcaption}
\usepackage{physics}
\usepackage{float}
\usepackage[normalem]{ulem}
\usepackage{comment}

\newcommand{\bnew}{\boldsymbol{B}}
\newcommand{\vnew}{\boldsymbol{v}}

\newcommand{\bturbind}{\bnew^{\prime}}

\newcommand{\vturbind}{\vnew^{\prime}}

\newcommand{\mean}[1]{\overline{#1}}

\newcommand{\vect}[1]{\boldsymbol{#1}}

\title[A new sub-grid model for MHD turbulence. II.]{Assessment of a new sub-grid model for magnetohydrodynamical turbulence. II. Kelvin-Helmholtz instability.}

\author[M. Miravet-Tenés, P. Cerd\'a-Dur\'an, M. Obergaulinger, and J.A.~Font]{
Miquel Miravet-Tenés,$^{1}$\thanks{E-mail: miquel.miravet@uv.es}
Pablo Cerdá-Durán,$^{1,2}$
Martin Obergaulinger$^{1}$,
and José A. Font$^{1,2}$
\\
$^{1}$Departament d'Astronomia i Astrofísica, Universitat de València, C/ Dr Moliner 50, 46100, Burjassot (València), Spain\\
$^{2}$Observatori Astronòmic, Universitat de València, C/ Catedrático José Beltrán 2, 46980, Paterna (València), Spain
}

\date{Accepted XXX. Received YYY; in original form ZZZ}

\pubyear{2023}

\begin{document}
\label{firstpage}
\pagerange{\pageref{firstpage}--\pageref{lastpage}}
\maketitle

\begin{abstract}
The modelling of astrophysical systems such as binary neutron star mergers or the formation of magnetars from the collapse of massive stars involves the numerical evolution of magnetised fluids at extremely large Reynolds numbers. This is a major challenge for (unresolved) direct numerical simulations which may struggle to resolve highly dynamical features as, e.g.~turbulence, magnetic field amplification, or the transport of angular momentum. Sub-grid models offer a means to overcome those difficulties. In a recent paper we presented {MInIT}, an MHD-instability-induced-turbulence mean-field, sub-grid model based on the modelling of the turbulent (Maxwell, Reynolds, and Faraday) stress tensors. While in our previous work MInIT was assessed within the framework of the magnetorotational instability, in this paper we further evaluate the model in the context of the Kelvin-Helmholtz instability (KHI). The main difference with other sub-grid models (as e.g.~the alpha-viscosity model or the gradient model) is that in {MInIT} we track independently the turbulent energy density at sub-grid scales, which is used, via a simple closure relation, to compute the different turbulent stresses relevant for the dynamics. The free coefficients of the model are calibrated using well resolved box simulations of magnetic turbulence generated by the KHI. We test the model against these simulations and show that it yields order-of-magnitude accurate predictions for the evolution of the turbulent Reynolds and Maxwell stresses. 
\end{abstract}

\begin{keywords}
{(magnetohydrodynamics) MHD -- turbulence -- instabilities -- methods: numerical}
\end{keywords}



\section{Introduction}

During the last few decades simulations of binary neutron star (BNS) mergers (see e.g.~\cite{Baiotti:2017,doi:10.1146/annurev-nucl-101918-023625} for reviews) have proven fundamental to advance our understanding of their dynamics and in assessing the role they play across various fields: relativistic astrophysics (BNS mergers being regarded as the central engine of short gamma-ray bursts~\citep{1999ApJ...524..262M} and kilonovae~\citep{doi:10.1146/annurev-nucl-102115-044819,1989Nature_nucleo}), gravitational physics (as prime sources of gravitational waves (GWs)~\citep{GW170817}), cosmology (as standard sirens~\citep{Nissanke:2010,Abbott:2017hubble}), and nuclear physics (to constraint the equation of state of dense matter at supranuclear densities~\citep{Read:2009,Ruiz:2017due,REzzolla:2018,Chatz:2020}).

BNS merger simulations are challenged by the inherent complexity of the problem, involving extreme physics. Arguably, one of the most complicated aspects to resolve in a simulation is the turbulent amplification of the magnetic field, a situation that spans the entire process, from the late inspiral to the post-merger phase, and involves different effects. In the merger phase and during the early post-merger magnetic fields are amplified via the Kelvin-Helmholtz instability (KHI) \citep{Obergaulinger:2010,producing_magnetar_fields_giacomazzo,eff_mag_field_kiuchi_cerda,Kiuchi:2018,Vigano:2020,Palenzuela:2022b,Aguilera-Miret:2023}, which leads to turbulent magnetic field values of ${\cal O}(10^{16})$ G. On longer timescales it has been argued that turbulent dynamos driven by the magnetorotational instability (MRI), may generate a large-scale magnetic field~\citep{Balbus:1991,Balbus:1998}. MRI-driven magnetohydrodynamical turbulence is also responsible for angular momentum transport in the remnant. Resolving these instabilities in the simulations is essential to properly capture the merger dynamics and late-time fate of the remnant, which in turn may potentially affect observables such as the multimessenger emission of the system. However, the small spatial scale at which the magnetic field forms at merger, ${\cal O}(10^{-2})$ m \citep{Guilet:2017}, severely limits current numerical approaches. Simulations with increasing resolution have gradually been reported \citep{2008PhRvD..78h4033B,2014CQGra..31g5012R,eff_mag_field_kiuchi_cerda,Kiuchi:2018,Kiuchi:2022,Murguia-Berthier:2021}, concluding that a spatial grid resolution of ${\cal O}(10)$ m is needed for the KHI to efficiently amplify the magnetic field. Those simulations are computationally extremely expensive, involving $\sim {\cal O}(10 \, \, \mathrm{million})$ CPU hours, which renders unfeasible  a systematic study of the magnetorotational evolution of BNS merger remnants. 

Attempts to overcome those limitations rely on sub-grid models to describe the small-scale dynamics \citep{smagorinsky,Leonard:1975,mueller:2002,ogilvie,Pessah:2006,Radice:2017,Radice:2020}. These models have recently begun to being used  in the context of BNS mergers~\citep{producing_magnetar_fields_giacomazzo,general_rel_kiuchi_shibata,Shibata:2021,extension_subgrid_vigano,paper_gradient_test,Vigano:2020,Palenzuela:2022,Aguilera-Miret:2022}. In a recent paper~\citep{Miravet:2022} we joined these efforts by  presenting a new sub-grid model based on the proportionality relations between the components of the turbulent stress tensors and the evolution of a turbulent kinetic energy density. This new model, dubbed {MInIT} (for {\it MHD-instability-induced-turbulence}), was assessed in \cite{Miravet:2022} in the particular case of the MRI. It was found that {MInIT} captures the development of turbulent stresses even if the numerical resolution of the MRI simulations is
not sufficient. In the current work we continue the study initiated in \cite{Miravet:2022} by assessing {MInIT} in the context of the KHI. 
The results reported in this paper indicate that our sub-grid model performs as satisfactorily as its MRI version, yielding  order-of-magnitude accurate predictions of the Maxwell and Reynolds turbulent stresses attained in box numerical simulations of the KHI. Together with the findings of~\cite{Miravet:2022} the present investigation  increases our confidence in our new sub-grid model before we develop a general-relativistic formulation that can be employed for actual BNS merger simulations, which are envisaged in our short-term plan. 

The paper is organized as follows: in Section~\ref{sect: sec2} we discuss the mean-field formalism used to separate numerically resolved quantities from the small-scale turbulent ones. Next, in Section~\ref{sect: sec3} we show the basis of the MInIT model applied to the KHI. In Section~\ref{sect: sec4} we describe the box numerical simulations of the KHI we use to carry out our testing of the sub-grid model. The results of the various tests are also reported in this section. Finally, our conclusions are summarized in Section~\ref{sect: sec5}. 

\section{Mean-field magnetohydrodynamics}
\label{sect: sec2}

We start by briefly reviewing the Newtonian ideal MHD equations which form the mathematical framework for our study. These equations couple the different variables of a plasma, like the gas pressure, the mass density, the velocity and the magnetic fields. We can express the system of MHD equations in flux-conservative form
\begin{equation}\label{mhd_eqs}
   \partial_t\boldsymbol{C}+\partial_j\boldsymbol{F}^j = 0\,, 
\end{equation}
where index $j$ runs across all possible values of the spatial coordinates and Einstein's summation convention is assumed. In this equation  $\boldsymbol{C}$ is the vector of conserved variables
\begin{equation}\label{state_vect}
  \boldsymbol{C} =   \left[\begin{array}{c}
         \rho  \\
        N^i \\
        U \\
        B^i 
        \end{array} \right],
\end{equation}
comprising the mass density, the momentum density, the energy density, and the magnetic field, respectively. Moreover, the flux vector in the spatial direction $j$, $\boldsymbol{F}^j$, reads
\begin{equation}\label{flux_vect}
  \boldsymbol{F}^j \equiv  \left[\begin{array}{c}
         \rho v^j \\
         \rho v^iv^j-B^iB^j+\delta^{ij}\big[p+B^2/2\big] \\
        v^j\big[U+p+B^2/2\big]-(v_kB^k)B^j \\
        v^jB^i-v^iB^j 
        \end{array} \right],
\end{equation}
where $p$ is the gas pressure, $v^i$ is the velocity, and $B^2 = B_i B^i$. 

Unresolved, sub-grid-scale terms can appear in the previous system of equations through the application of the mean-field MHD formalism \citep{krause}. Given any field $\boldsymbol{A}$, it can be decomposed into the sum of an averaged, resolved part, $\overline{\boldsymbol{A}}$, and a turbulent, unresolved contribution, $\boldsymbol{A}^{\prime}$. The first term corresponds to the expectation value of the field, and the average can be both spatial or temporal. By following \cite{book}, the filtering operator can be defined as either
\begin{equation}
    \overline{\boldsymbol{A}} = \frac{1}{V}\int_{V} \boldsymbol{A}(t,\boldsymbol{x})\, d^3\boldsymbol{x}\,,
\end{equation}
for a spatial average in a scale of order $\lambda$ (having thus a volume $V \propto \lambda^3$), or
\begin{equation}
	\overline{\boldsymbol{A}} = \frac{1}{\tau}\int_{\tau} \boldsymbol{A}(t,\boldsymbol{x}) \, dt \,,
\end{equation}
for a time average in a timescale $\tau$. 

Since the average of the fluctuating part is zero, the only possible terms related to the fluctuating part in the mean-field equations are the mean of combinations of products of two or more fluctuating variables. As in \cite{Miravet:2022} we only consider fluctuations of the velocity ($\vturbind$) and of the magnetic field ($\bturbind$). With these considerations, the fluctuating part in the equations can be written in terms of the following tensors:
\begin{subequations}\label{stress_tensors}
 \begin{align}
	M_{ij} & = B^{\prime}_i B^{\prime}_j,\\
	R_{ij} & = v^{\prime}_i v^{\prime}_j, \\
	F_{ij} & = v^{\prime}_i B^{\prime}_j-v^{\prime}_j B^{\prime}_i\,, 
\end{align}
\end{subequations}
namely, the Maxwell, Reynolds and Faraday stress tensors, respectively, as defined in \cite{ogilvie}. When expressing, and then averaging, the MHD equations in terms of resolved and unresolved quantities, the system takes the form
\begin{equation}\label{mf_mhd_eqs}
   \partial_t\widetilde{\boldsymbol{C}}+\partial_j\widetilde{\boldsymbol{F}}^j = \boldsymbol{S}(\overline{R}_{ij},\overline{M}_{ij},\overline{F}_{ij})\,, 
\end{equation}
where quantities with a tilde refer to identical functional expressions as in Eqs.~\eqref{state_vect} and \eqref{flux_vect} but for the mean quantities $\overline{\boldsymbol{B}}$ and $\overline{\boldsymbol{v}}$ (instead of $\boldsymbol{B}$ and $\boldsymbol{v}$). Note that $\widetilde{\boldsymbol{F}}^0 \ne \overline{\boldsymbol{F}}^0$ and $\widetilde{\boldsymbol{F}}^j \ne \overline{\boldsymbol{F}}^j$, and the difference between the terms with tilde and their average is what appears in the source term, $\boldsymbol{S}$, which is a function of the mean stresses. 

\section{MHD-instability-induced-turbulence (MInIT) mean-field model for KHI}
\label{sect: sec3}

Following \cite{Miravet:2022} we aim at prescribing an evolution equation for the energy density stored in KHI-induced turbulence, that can be used to estimate the different stresses appearing in the mean-field Eqs.~\eqref{mf_mhd_eqs}. The drivers of the KHI are large-scale shears present in the bulk motion of the fluid (in the case of BNS mergers this happens particularly during the merger) that induce a turbulent cascade. The main difference with the MRI is that in that case turbulence is driven by shear flows occurring at small unresolved scales. Therefore, to model the KHI we do not worry about tracking the energy in the shear itself (as we did in \cite{Miravet:2022}) but only in the turbulent part. This results in a model that is somewhat simpler than the MRI model described in \cite{Miravet:2022}. 

To build the KHI MInIT model we need three ingredients: the growth rates of the KHI computed in terms of the averaged quantities, an evolution equation for the turbulent energy density, and a closure model to compute the stresses in terms of this energy density. We address this three items next.

\subsection{KHI growth rates}
\label{sec:growthrates}

The growth rate of the KHI in a magnetized fluid was studied by \cite{miura_1982} as a function of the sound Mach number, $M_{\rm s} = v/c_{\rm s}$, the Alfvén Mach number, $M_{\rm A} = v/v_{\rm A}$, and the relative orientation of the velocity and the magnetic field. In the case of a magnetic field parallel to the velocity field, the fastest growing mode appears at the limit of zero magnetic field ($M_{\rm A}^{-1} = 0$). For a flow with $M_{\rm s}=1$, the fastest growing mode has a wavenumber (parallel to the velocity) and growth rate
\begin{equation} \label{growth_rate_khi}
k_{\rm KH}\approx0.4/a_l \, , \qquad 
\gamma_{\rm KH} \approx 0.14\frac{v}{a_l}\,,
\end{equation}
respectively, where $2v$ here is the variation of the velocity across the shear and $a_l$ is the characteristic scale length of the variation of the velocity in this region. The wavelength of this unstable mode is thus
\begin{equation}
\lambda_{\rm KH} = \frac{2\pi}{k_{\rm KH}} \approx 15.8 a_l.
\label{eq:wavelength:khi}
\end{equation}

For higher magnetic fields both the normalized growth rate ($\gamma_{\rm KH} \frac{a_l}{v}$) and the wavenumber decrease, but as long as the magnetic field is subdominant ($M_{\rm A}>1$), both quantities do not differ much from the maximum values.
The normalized growth rate  increases with decreasing sound Mach number $M_{\rm s}$ (about a $20\%$ increase for half the sound Mach number), and increases with increasing Alfvén Mach number $M_{\rm A}$. 

As noted by \cite{miura_1982} similar conclusions hold if the magnetic field is transversal to the velocity field. In particular, the growth rates for the transversal case are not larger than for the parallel case when comparing their values for the same Mach numbers. The shape of the shear used for the analysis may also affect the exact values of the growth rates \citep[see][for the unmagnetized case]{chandra_1961} but not their order-of-magnitude. Therefore, for a fluid with a non-dominant magnetic field and $M_{\rm s} \sim 1$ we expect the above numbers to be adequate order-of-magnitude estimates. This is the case for the numerical setup used in the simulations in this work.

\subsection{Evolution equation of the KH energy density}
\label{sec:evolution:e}

We define the turbulent kinetic energy density in the KHI as 
\begin{equation}
    e_{\rm KH} =
    \frac{1}{2} \, \rho \, \sum_{i=1}^{3}\overline{R}_{ii} \,. \label{eq:eKH}
\end{equation}
Following \citet{Miravet:2022} we can write the evolution equation for this turbulent energy density as
\begin{equation}
    \partial_t e_{\rm KH }+ \partial_i (\overline{v}_i\,e_{\rm KH}) = S^{\rm (KH)}\,,
\end{equation} 
where $S^{\rm (KH)}$ is a source term that acts as a generator and sink of the turbulent energy density. As in our previous work with the MRI, we neglect the stretching terms, which are irrelevant and add unnecessary complication to this work, and would also make the equations non-conservative (see~\cite{Miravet:2022} for more details). 

A closed and simple form for the source terms can be justified taking into account some information we derived in the last few subsections, namely: i) the KH channel flows grow exponentially with a growth rate $\gamma_{\rm KH}$ and therefore $e_{\rm KH}$ will grow with $2\gamma_{\rm KH}$; ii) the vortices from the turbulence are broken into smaller ones in a turbulent cascade until dissipation occurs at the physical dissipation scale, where the small scale, kinetic and magnetic energy transforms into thermal energy \citep[see e.g.][chapter III]{LANDAU198795}. Considering these points, the evolution equation for the KH energy density becomes
\begin{equation}\label{energy_dens_ev}
    \partial_t e_{\rm KH} +\partial_i(\mean{v}_i e_{\rm KH}) = 2\,\gamma_{\rm KH}\,e_{\rm KH}-S_{\rm TD} \,,
\end{equation}
where $\gamma_{\rm KH}$ is the growth rate discussed in the previous section. This equation shows the energy flow in a KH-unstable system. The first term in the RHS, $2 \, \gamma_{\rm KH} e_{\rm KH}$ takes energy from the large-scale quantities, which increases the turbulent energy. The quantity $S_{\rm TD}$ represents the turbulent energy dissipation at the end of the Kolgomorov scale, which dissipates the kinetic (and magnetic) energy into thermal (internal) energy, and is transferred back to the large-scale quantities.

Concerning the Kolgomorov term, we are using the same expression  obtained in \cite{Miravet:2022}, which was proposed following the arguments in Section 33 of \cite{LANDAU198795}, namely
\begin{equation}\label{kolgomorov}
    S_{\rm TD} = C \frac{e_{\rm KH}^{3/2}}{\sqrt{\rho}\lambda_{\rm KH}}\,,
\end{equation}
where $C$ is a constant that can be calibrated from numerical simulations (see next sections).

In a practical application of the MInIT model one would have a sheared velocity profile that is covered by a number of grid points and one would want to model the sub-grid scale. In practice this means that, depending on the resolution, the KHI could already be partially resolved. However, it is unlikely
that the full turbulent cascade is resolved by the simulation, given the extremely large Reynolds numbers of the astrophysical applications considered (e.g.~BNS mergers). Therefore, the shear leading to the sub-grid turbulence can be either large-scale shear present in the simulations (e.g. the shear produced at the merger of two neutron stars), or that generated by eddies of the  resolved turbulence of the simulation (e.g. large-scale vortexes produced by KHI in the aforementioned large-scale shear). One characteristic of Kolgomorov-type turbulence, such as the one produced by the KHI (at least in the unmagnetized case) is that the energy flux towards lower scales is constant across wave numbers and proportional to $v^3/a_l$ \citep{LANDAU198795}, where $v$ is here the velocity of the vortex at each scale $a_l$. One can consider that vortices at each scale break up into smaller-scale vortices due to KHI that arises from the shear produced by those vortices. One can estimate the scaling of the vortex velocity, wavelength, and growth rate of the KHI as
\begin{equation}
 v \propto (a_l)^{1/3} \, , \qquad \lambda_{\rm KH} \propto a_l\, , \qquad \gamma_{\rm KH} \propto (a_l)^{-2/3}.
 \label{eq:gammaScaling}
\end{equation}
\begin{table}
    \centering
\begin{tabular}{|ccc|}
\hline
    Name & $\mean{B}_{x0}$ &  $N_{\rm cells}^3$ \\
    \hline
    \hline
    KH-L1 & $3\times 10^{-4}$ & $128^3$  \\
    KH-L2 & $1\times 10^{-3}$ & $128^3$  \\
    KH-L3 & $3\times 10^{-2}$ & $128^3$  \\
    \hline
    KH-M1 & $3\times 10^{-4}$ & $256^3$  \\
    KH-M3 & $3\times 10^{-2}$ & $256^3$  \\
    \hline
    KH-H1 & $3\times 10^{-4}$ & $512^3$  \\
    KH-H2 & $1\times 10^{-3}$ & $512^3$  \\
    KH-H3 & $3\times 10^{-2}$ & $512^3$  \\
\hline
\end{tabular}
    \caption{List of KHI simulations. The middle column depicts the initial value of the magnetic field for all our KHI box simulations. The number of cells for the low, medium, and high resolution simulations is reported in the last column. The size of each side of the box is $L = 1$ and the characteristic width of the shear layer is $a_l = 0.01$.}
    \label{tab:table_sim}
\end{table}
Since by increasing the grid resolution of the simulation one resolves smaller scales $a_l$, the KHI at sub-grid scales has larger growth rates but, at the same time, $S_{\rm TD}$ becomes larger as well. The amount of energy density in turbulence in the fully turbulent case can be estimated by making the r.h.s of Eq.~\eqref{energy_dens_ev} equal to zero, which implies a zero growth of the instability. This yields an estimated value of
\begin{equation}
e_\textrm{ KH, fully turbulent} \propto (a_l)^{2/3}.
\label{eq:EneFullTurbulence}
\end{equation}
The implication is that increasing grid resolution leads to a smaller amount of turbulent energy density in the sub-grid model, as one would expect. This scaling relation can be tested in numerical simulations as shown below in Section~\ref{sec:filter}. We also note, however, that as the resolution is increased the differential equations to be solved become increasingly stiff, from the numerical point of view.

\begin{figure*}
    \centering
    \includegraphics[width=\textwidth]{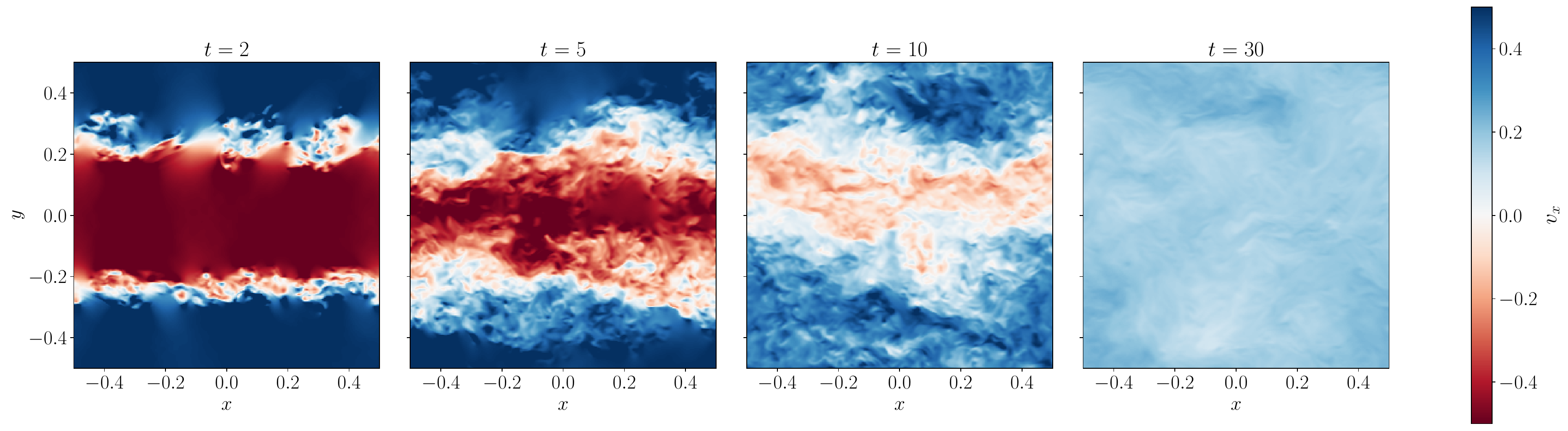}
    \caption{Slices of the $x$-component of the velocity field at the $xy$-plane for the simulation KH-M1. The panels show different times of the simulation, $t={2,5,10,30}$. The KHI sets in from a shear layer configuration and reaches an isotropic and turbulent configuration at late times ($t=30$).}
    \label{fig:vx_cuts}
\end{figure*}

\begin{figure*}
    \centering
    \includegraphics[width=\textwidth]{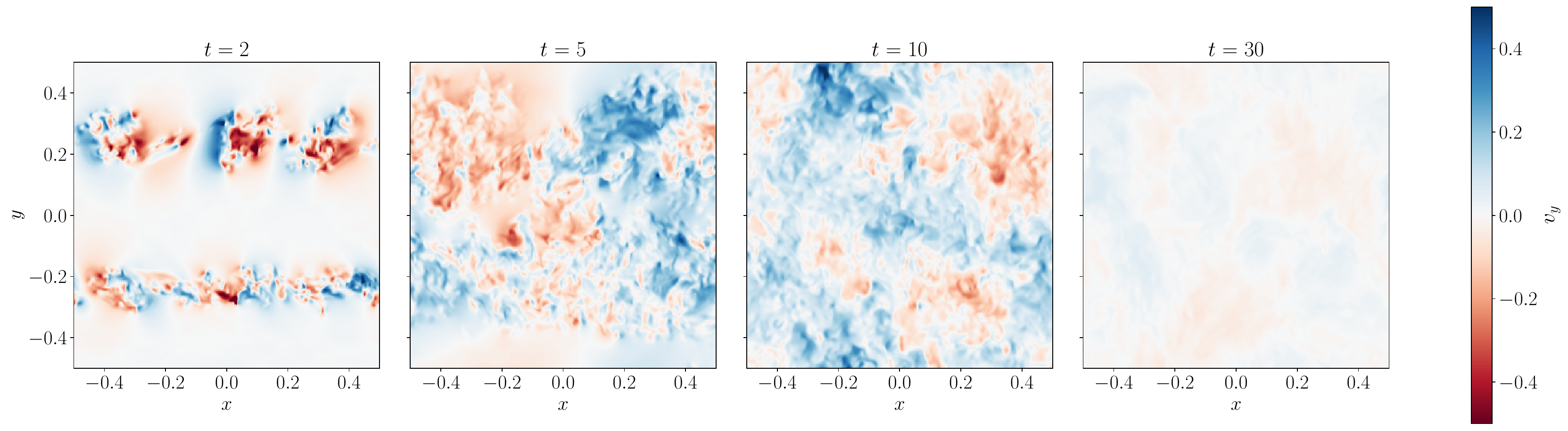}
    \caption{Slices of the $y$-component of the velocity field at the $xy$-plane for the simulation KH-M1. From left to right the selected times correspond to  $t={2,5,10,30}$.}
    \label{fig:vy_cuts}
\end{figure*}

\subsection{Closure relation}
\label{sec:closure}

The main assumptions of the closure relation used for the MInIT model in \citet{Miravet:2022} is that the different stress tensors are proportional to the  turbulent kinetic energy density and that the proportionality coefficients do not depend explicitly on time and position, which implies that the stress tensors have the same time dependence as $e_{\rm KH}$. The closure relation between the turbulent energy density, $e_{\rm KH}$ and the stresses is the following:
\begin{subequations}\label{closure_coeffs}
\begin{align}
	\mean{M}_{ij}(t,\vect{r}) & = \alpha^{\rm KH}_{ij} \,e_{\rm KH}(t,\vect{r})\,, \label{Maxwell_coeff}\\
	\mean{R}_{ij}(t,\vect{r}) & = \frac{1}{\mean{\rho}(t, \vect{r})}  \beta^{\rm KH}_{ij} \,e_{\rm KH}(t,\vect{r})\,, \\
	\mean{F}_{ij}(t,\vect{r}) & = \frac{\gamma^{\rm KH}_{ij}}{\sqrt{\mean{\rho}(t,\vect{r})}} e_{\rm KH}(t,\vect{r}) \,, 
\end{align}
\end{subequations}
 where the factors involving the mass density in the last two equations are added in order to make the coefficients dimensionless. These coefficients are calibrated using numerical simulations and their values are discussed in the next sections.

\section{Results}
\label{sect: sec4}

\subsection{KHI box simulations}
\label{sec:boxsim}

\begin{figure*}
    \centering
    \includegraphics[width=0.9\linewidth]{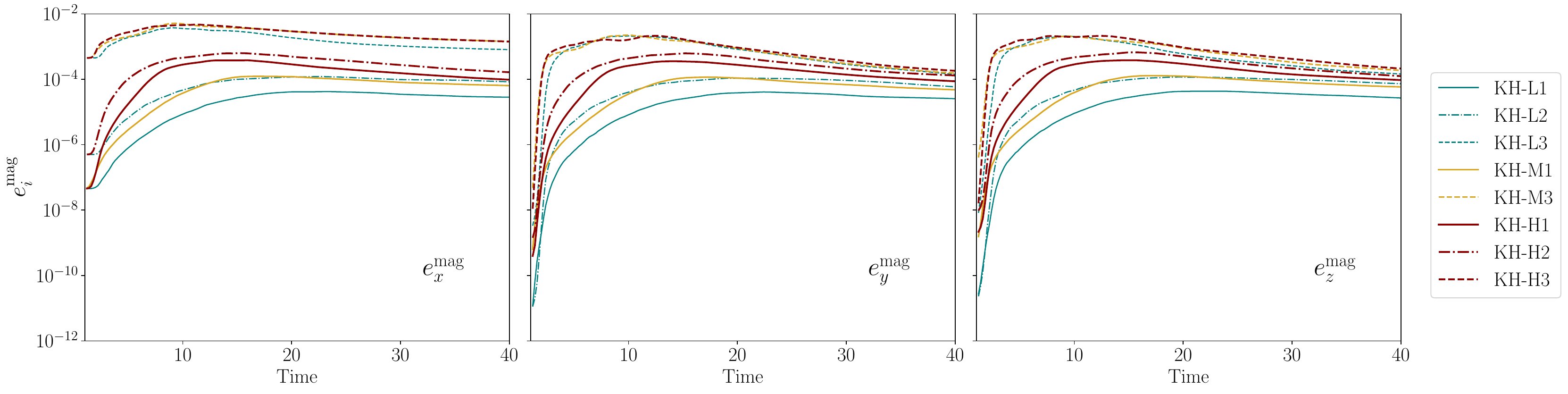}
    \caption{Time evolution of the $x$, $y$, and $z$ components of the averaged magnetic energy density. Curves with different colors correspond to different grid resolutions, as indicated by the legend on the right (see also Table~\ref{tab:table_sim}).  Solid, dash-dotted, and dashed lines correspond to $\overline{B}_{x0} = 3\times 10^{-4}$, $1\times 10^{-3}$ and $3\times 10^{-2}$, respectively.}
    \label{fig:mag_energy}
\end{figure*}

\begin{figure*}
    \centering
    \includegraphics[width=0.9\linewidth]{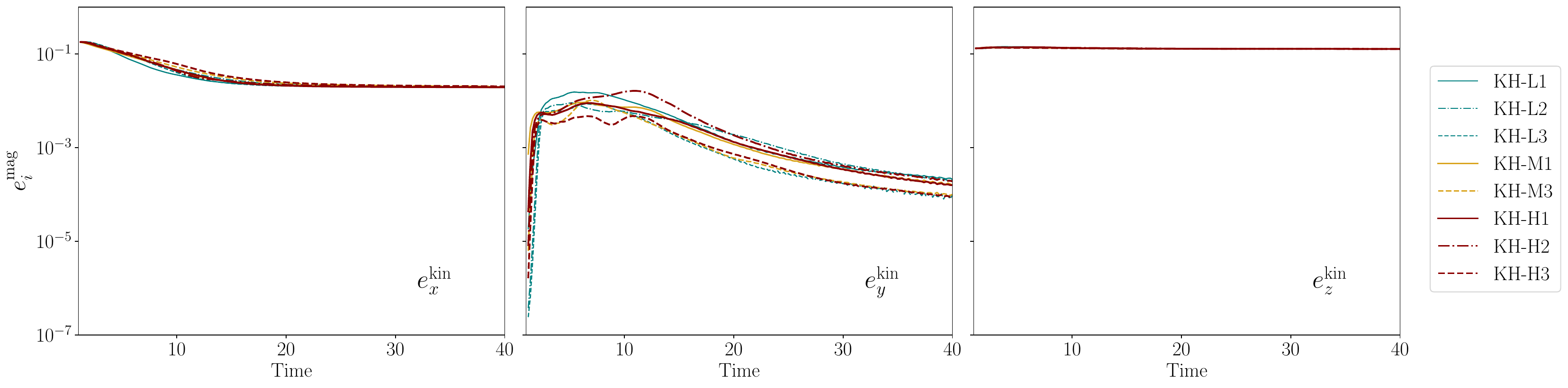}
    \caption{Time evolution of the $x$, $y$, and $z$ components of the averaged kinetic energy density. As in Fig.~\ref{fig:mag_energy}, the colors represent different gird resolutions and the various linestyles correspond to different initial magnetic field strengths $B_{x0}$.}
    \label{fig:kin_energy}
\end{figure*}

To calibrate and test the performance of the model we perform a series of box simulations of the KHI displaying turbulence. We consider a periodic (in all directions) 3D Cartesian box of size $L^3$, with $L=1$. The initial fields are similar to the ones employed by \cite{paper_gradient_test}:
\begin{subequations}
\begin{align}
    \rho & = \rho_0+\rho_1\tanh{\Bigg(\frac{|y|-y_{\pm}}{a_l}\Bigg)}\,,  \\
     v_{x} & = v_{x0}\,{\rm{sgn}}(y)\tanh{\Bigg(\frac{|y|-y_{\pm}}{a_l}\Bigg)}+\delta v_{x}\,, \\
     v_y & = \delta v_y\,{\rm{sgn}}(y)\exp{\frac{(|y|-y_{\pm})^2}{\sigma_y^2}}\,, \\
     v_z & =  v_{z0}\,{\rm{sgn}}(y)\exp{\frac{(|y|-y_{\pm})^2}{\sigma_z^2}}+\delta v_z\,, \\
     B_x & = \overline{B}_{x0},\hspace{10pt} B_y = 0,\hspace{10pt} B_z = 0 \,, \\
     p & = p_0 \,.
\end{align}
\end{subequations}
These initial conditions describe two shear layers located at $y = y_{\pm}$. The values considered in our simulations are $y_{\pm} = \pm 0.25$, $\rho_0 = 1.5$, $\rho_1 = 0.5$, $v_{x0}=v_{z0}=0.5$, $p_0 = 1$ and $a_l = 0.01$, where $a_l$ is the characteristic width of the shear layer. The parameters $\sigma_y^2 = 0.01$ and $\sigma_z^2 = 0.1$ are the scale of the initial velicity perturbation in the $y$-direction and the profile of $v_z$, respectively. The value of $\overline{B}_{x0}$ is a free parameter that we change for the different simulations. The specific values are reported in Table~\ref{tab:table_sim}. The sound Mach number, $M_s = v_0/c_s$, is fixed to unity in all the simulations. The initial data are perturbed by seeding small random perturbations in velocity, $\delta v_i$, with amplitudes of $10^{-3}$. We note that this is different to the sinusoidal perturbations used in \cite{paper_gradient_test}. We consider an ideal gas equation of state,
\begin{equation}\label{eos}
    p = (\Gamma-1)\rho \epsilon\,,
\end{equation}
with $\Gamma = 4/3$. 

We evolve these initial conditions using the \textsc{Aenus} code \citep{Obergaulinger-2008} which solves the ideal MHD equations in its conservative form using finite-volume methods. All the simulations were performed using the HLL flux formula, a monotonicity-preserving (MP) reconstruction of $5^{\mathrm{th}}$ order and a $3^{\mathrm{rd}}$ order Runge-Kutta time integrator.

The general qualitative behaviour of the simulations can be seen in Figure~\ref{fig:vx_cuts}. This figure displays a selection of slices in the $xy$ plane of the $x$-component of the velocity field, $v_x$, for the simulation KH-M1 at four different times. At the beginning of the simulation there are two shear layers which separate regions with $v_x \approx \pm 0.5$. Once the instability sets in, vortices develop on these shears and a turbulent isotropic state is reached. The process progressively smooths and finally erases the initial shear by transporting linear momentum in the $y$ direction. Figure~\ref{fig:vy_cuts} shows similar slices at the same snapshots but for the $y$-component of the velocity. It is apparent that at early times $v_y$ is characterized by localized perturbations that later merge into {a more extended region} from $t=5$ onwards. 

Figure~\ref{fig:mag_energy} shows the time evolution of the contribution of each magnetic-field component to the total averaged magnetic energy density 
\begin{equation}
    \overline{e}^{\rm mag}_i = \frac{\overline{B_i^2}}{2}\,,
\end{equation}
for different grid resolutions and initial magnetic field strength $\overline{B}_{x0}$. The saturation level of the turbulent magnetic field depends on the two of them.  At low resolution ($128^3$ cells, blue curves) we find a monotonic increase with the initial field strength (the three values used for the field strength can be distinguished by a different type of line, as indicated in the caption).  Keeping the initial field fixed, the energy of the final turbulent field increases when doubling the resolution to the intermediate grid of $256^3$ cells.  A subsequent twofold increase of the resolution to the finest grid of $512^3$ cells, however, does not change the results by the same degree.  Therefore, at this resolution, we  approach a state in which the differences between weak and strong initial fields are small.  For the case of model KH-H3, with the largest initial magnetic field of the models surveyed, $\overline{B}_{x0} = 3\times 10^{-2}$, the saturated magnetic energy in the $x$-direction is more than an order of magnitude larger than the energy of the rest of the simulations, with lower initial magnetic fields.

\begin{figure*}
    \centering
    \includegraphics[width=\linewidth]{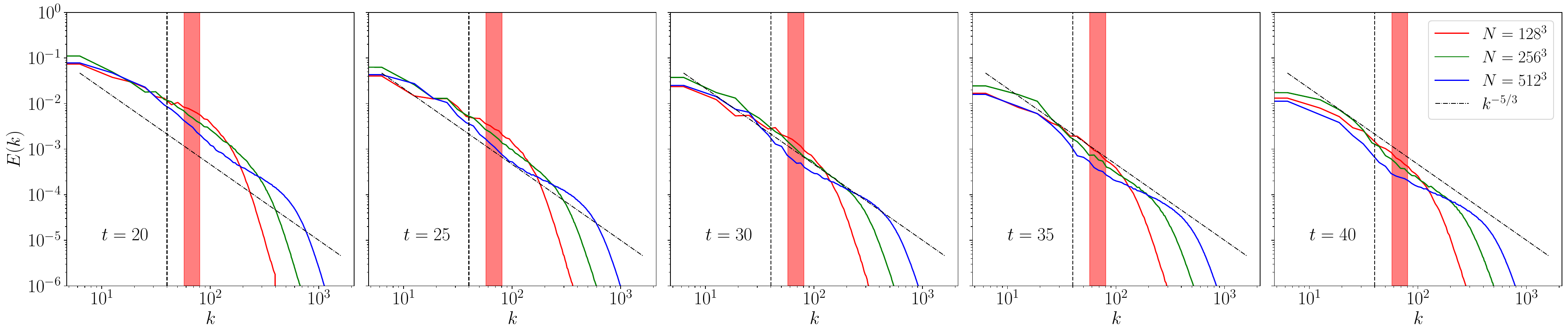}
    \caption{Snapshots of the kinetic energy spectra for different gird resolutions (corresponding to models KH-L1, KH-M1, and KH-H1 in Table~\ref{tab:table_sim}). The dashed line corresponds to the Kolgomorov slope $k^{-5/3}$. The red shaded region shows the range in which box filters are applied. All of them are located inside the inertial range of spatial scales. The vertical black line refers to the wavenumber of the initial shear, $k_{\rm KH}$.}
    \label{fig:kin_spectra}
\end{figure*}

Correspondingly, Figure~\ref{fig:kin_energy} shows the time evolution of the contribution of each component of the kinetic energy to the total averaged value of this quantity for the same simulations shown in Fig.~\ref{fig:mag_energy}. The expression for each component is 
\begin{equation}
    \overline{e}^{\rm kin}_i = \frac{1}{2}\overline{\rho v_i^2} \,.
\end{equation}
Leaving aside the small effect introduced by the initial random perturbations, the only non-vanishing components of the initial velocity field are $v_x$ and $v_z$ (left and right panels, respectively). These components remain nearly constant during the evolutions (and so do $e_x^{\rm kin}$ and $e_z^{\rm kin}$). However, the $e_y^{\rm kin}$ component (middle panel) grows exponentially during the early phase of the evolution to saturate and then slowly decrease. No constant state has yet been reached by the end of our simulations. This behaviour of $e_y^{\rm kin}$ is due to the transport of linear momentum across the shear in the $y$ direction. This leads to the mixing of the positive and negative parts of the velocity field which results in the decrease of the velocity components shown in Fig.~\ref{fig:vy_cuts}, while conserving the initially (almost) zero linear momentum. No remarkable qualitative differences are found among the simulations. It is worth stressing that the $y$-component of the kinetic energy achieves a lower value for stronger initial magnetic fields, regardless of the resolution.

In Figure~\ref{fig:kin_spectra} we depict five snapshots of the evolution of the kinetic energy spectra for different grid resolutions. The spectra were calculated in the same way as in \cite{Simon:2009}. One can see that the inertial range of the spatial scales, parallel to the Kolmogorov slope $\propto k^{-5/3}$ (dashed line), goes from scales smaller than the characteristic scale of the initial shear (vertical line) to scales at which there is dissipation into internal energy. The red shaded region represents the interval of scales for which we employ box filters (see below), ensuring that we are inside the inertial range. As expected, as the grid resolution increases, the dissipation scales become smaller. 

\subsection{{Averaging procedure}}
\label{sec:filter}

Our aim is to compare the numerical box simulations with the MInIT model in order to calibrate its free coefficients. In this procedure the KHI box simulations play the role of the true solution of the ideal MHD equations (although it should always be kept in mind that the accuracy of the numerical solution does depend on the finite numerical resolution and on the order of the numerical method). Using the true solution we perform averages over a certain length scale in order to achieve a separation between the large-scale (averaged) features and the small (turbulent) scales that we want to test with MInIT. Given that the result of the simulation does not depend on the sub-grid model itself, this procedure provides an a-priori test of the model. On the other hand, a-posteriori tests would involve performing the simulations with the MInIT model coupled to the dynamics. Those go beyond the scope of the current paper and will be presented elsewhere.

The averaging procedure that we use consists in a box filter, identical to the one we employed in \cite{Miravet:2022}. We perform the average over $S_f^3$ cells, where $S_f = \Delta_f / \Delta$ labels each filter size, $\Delta$ is the size of the cell of the direct numerical simulation, and $\Delta_f$ is the filter size. Note that $\Delta_f$ is related to the scale $\lambda$ defined in Section~\ref{sect: sec2}. Following~\citet{paper_gradient_test} we characterize the box filter with the following normalized kernel, 
\begin{equation}\label{box_filter}
    F_i(|r_i-r^{\prime}_i|) = \left \{ \begin{array}{cc}
        1/\Delta_f & \mbox{ if } \,\, |r_i-r^{\prime}_i| \leq  \Delta_f^i/2 \,,\\
         0 & \mbox{ if } \,\, |r_i-r^{\prime}_i| > \Delta_f^i/2 \,,
    \end{array} \right.
\end{equation}
for each dimension $i$. The three-dimensional kernel is thus
\begin{equation}\label{box_3d}
    F (|\boldsymbol{r}-\boldsymbol{r}^{\prime}|) = \prod_i^3 F_i(|r_i-r^{\prime}_i|)\,.
\end{equation}

\begin{figure}
    \centering
    \includegraphics[width=\linewidth]{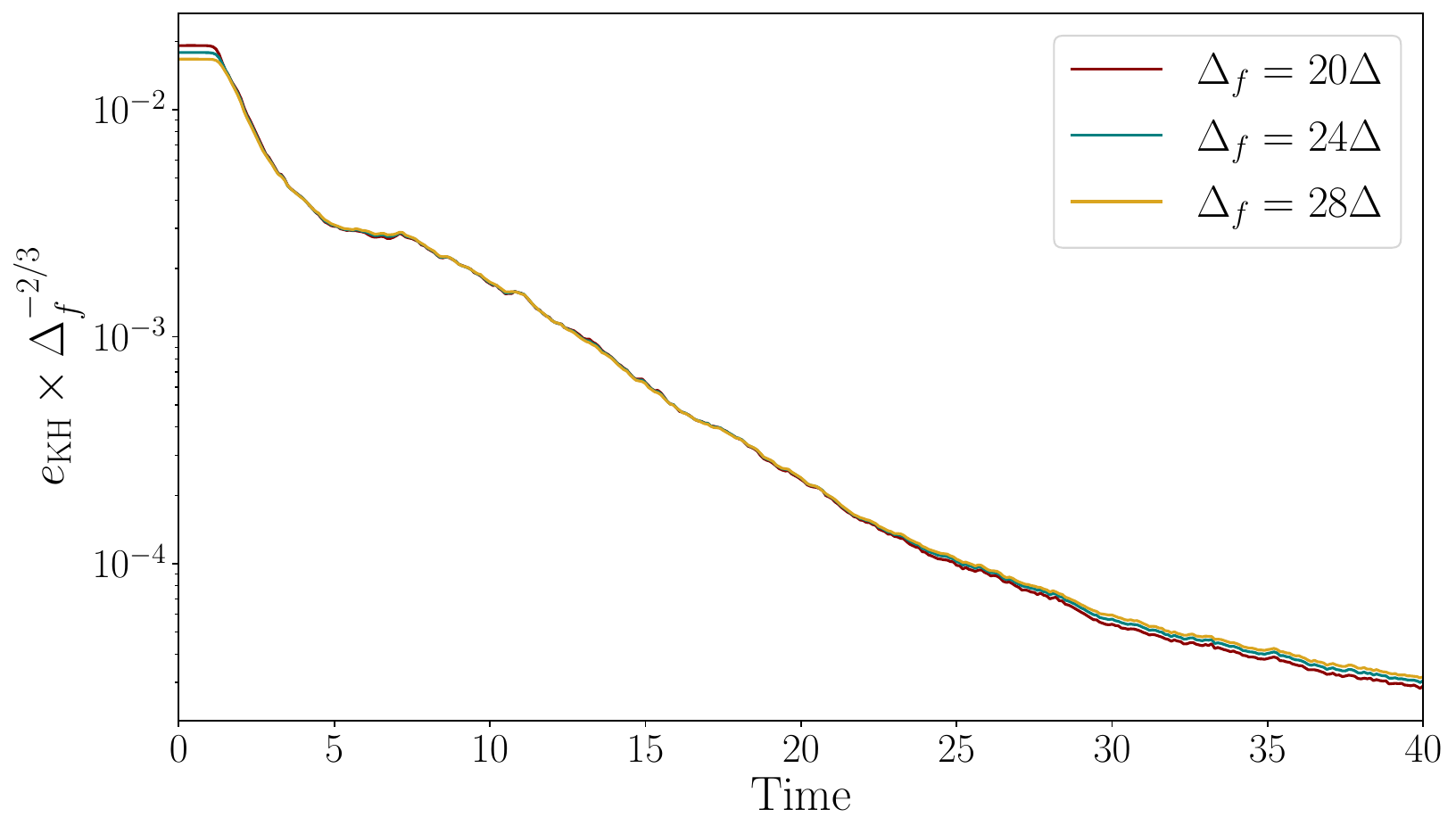}
    \caption{Time evolution of the quantity $e_{\rm KH} \Delta_f^{-2/3}$ for the simulation KH-M1 and for different averaging lengths $\Delta_f$.}
    \label{fig:e:scaling}
\end{figure}

\begin{figure}
    \centering
    \includegraphics[width=\linewidth]{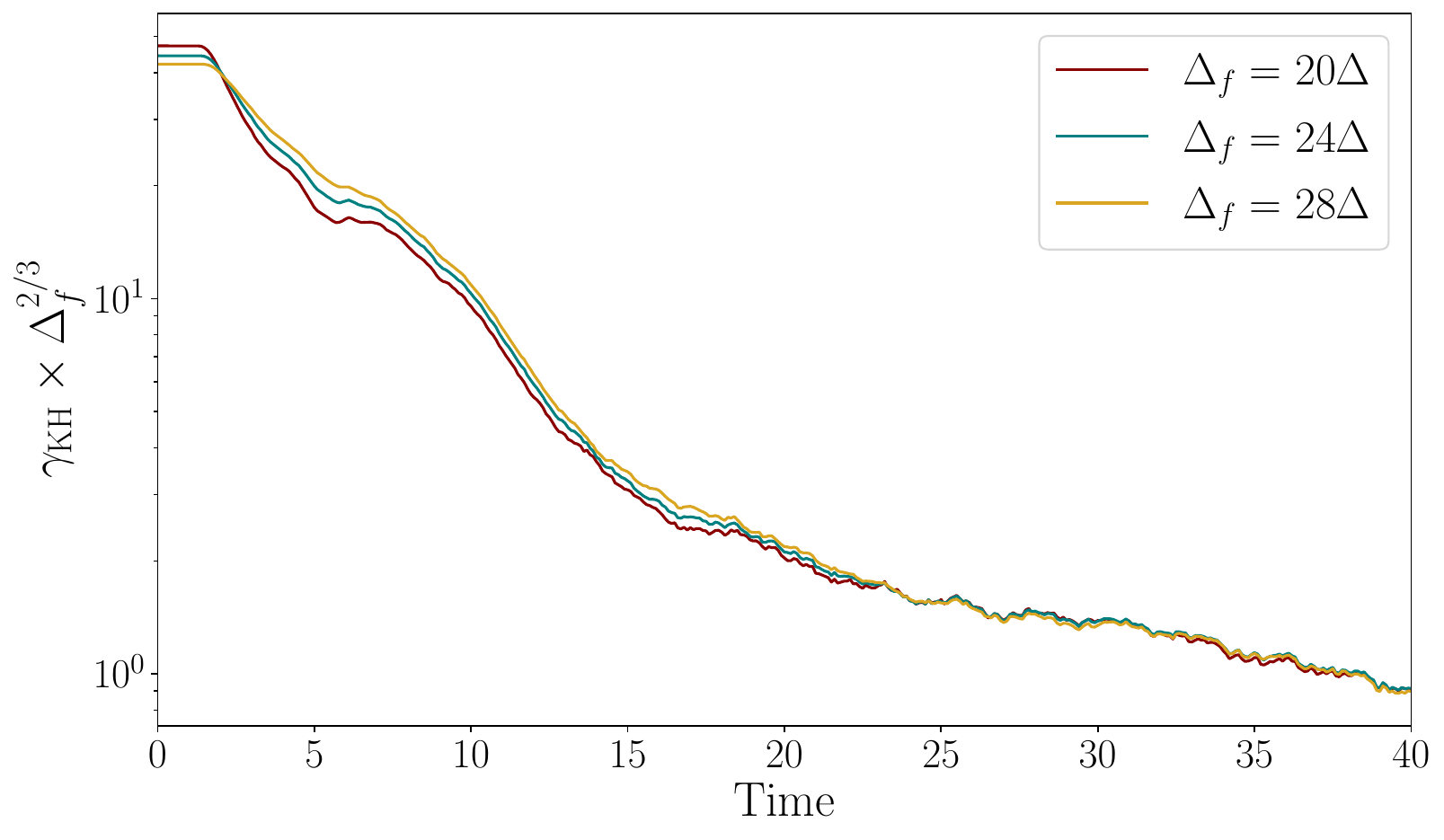}
    \caption{Same as Fig.~\ref{fig:e:scaling} but for quantity $\gamma_{\rm KH} \Delta_f^{2/3}$.}
    \label{fig:gamma:scaling}
\end{figure}

As a first application of this filter we test the validity of Eq.~\eqref{eq:EneFullTurbulence} that predicts that the kinetic energy density in the sub-grid scales for fully developed turbulence depends on the filter size $\Delta_f$. Figure~\ref{fig:e:scaling} shows the evolution of $e_{\rm KH} \Delta_f^{-2/3}$ during the entire evolution of simulation KH-M1 for three filter sizes. We observe that this quantity is almost independent of $\Delta_f$, even at early times. At late times ($t\gtrsim 20$) small deviations become more visible. Those could be related to the fact that, since the turbulent cascade is not fully resolved down to the dissipation scale, the smallest scale is only a factor $\sim 10$ smaller than the averaging scale. Therefore, the assumption that there is a large separation of  scales, used in the derivation of Eq.~\eqref{eq:EneFullTurbulence}, starts being compromised. This results in a small deficit in the turbulent energy density due to the unresolved scales. In addition, in Figure~\ref{fig:gamma:scaling} we depict the evolution of $\gamma_{\rm KH} \Delta_f^{2/3}$ for the same resolution and filter sizes. According to Eq.~\eqref{eq:gammaScaling} this quantity should also be independent of the filter size. As the figure shows, during the turbulent phase of the simulation ($t\gtrsim 20$) the scaling holds precisely.

\subsection{Numerical implementation and calibration of the MInIT model}

\subsubsection{Energy density evolution equations in the MInIT model}

In order to apply the MInIT model one has to integrate numerically Eqs.~\eqref{energy_dens_ev} in time, starting with appropriate initial values at $t=0$, $e_{\rm KH}(0)$. Those initial conditions are discussed in Section~\ref{sec:OPTC} below. For our KHI  simulations the average velocity $\overline{\boldsymbol{v}}$ is almost zero for the $x$ component, since both signs cancel out when  averaging over the entire box (and taking both sides of the shear layers). Likewise, the averaged component in the $y$ direction is also zero since we only consider random perturbations for this component. The only non-zero component will be $\overline{v}_z$, and we should therefore consider the advective term in this direction. However, since the simulations have periodic boundary conditions in the $z$ direction, any spatial average along this direction will be independent of the advection. Therefore, for the same reason as in \cite{Miravet:2022}, we can safely neglect the advection term for the tests in this paper. 

We perform our analysis in points centered at the shear layers of the initial conditions, $y_\pm = \pm 0.25$, points where we compute all averaged quantities. Performing the analysis in multiple points in the $xz$-plane (and different times) allow us to build more statistics for our results. Additionally, we compute auxiliary averages centered in $y_\pm \pm \Delta_f$, that allow us to evaluate $y$ derivatives at the shear. These different averages mimic the coarse numerical cells that one would have in simulations using the sub-grid model. They are used to compute the quantities needed in the model, in particular the growth rate and the KHI length scale needed for the turbulent source term given by Eq.~\eqref{kolgomorov}.

For the growth rate of the turbulent energy density, $\gamma_{\rm KH}$, we assume 
the values discussed in Section~\ref{sec:growthrates} obtained from \cite{miura_1982} (see Eq.~\eqref{growth_rate_khi}). To evaluate this rate we need information from the simulation about the value of the velocity jump of the shear and of the characteristic length scale $a_l$. Those values are known for the initial conditions but they are significantly modified as the simulation proceeds. Hence, a way of evaluating the rate is to substitute the term $v/a_l$ by $\partial_y v_x$ in Eq.~\eqref{growth_rate_khi}. Note that for the initial data this yields exactly $\partial_y v_x = v_{x0}/a_l$. Numerically, the partial derivative is evaluated using centered finite differences to the coarse grid cells defined above,
\begin{equation}\label{grate_num}
 \gamma_{\rm KH, \pm}^{\rm num} \approx A_{\gamma} \partial_y v_x \approx  A_{\gamma} \frac{\bar{v}_y (y_\pm + \Delta_f) -\bar{v}_y (y_\pm - \Delta_f) }{2\Delta_f},
\end{equation}
and then averaged for different points in the $xz$ plane and over the $\pm$ shears to obtain a single value of $\gamma_{\rm KH}$. Note that instead of using the same numerical constant as in Eq.~\eqref{growth_rate_khi} we use a generic constant $A_\gamma$. This constant should be close to $0.14$ according to \cite{miura_1982} but, since it may depend on the exact conditions at the shear (see Section~\ref{sec:growthrates}) we leave it free at first; its value is discussed in the next sections.

Regarding the Kolgomorov term from Eq.~\eqref{energy_dens_ev}, the wavelength of the instability is discussed in Section~\ref{sec:growthrates} following the work of \cite{miura_1982} (see Eq.~\eqref{eq:wavelength:khi}), and is proportional to the shear length scale $a_l$. Since we only want to model shears at the lower possible scale (see discussion in Section~\ref{sec:evolution:e}) we assimilate the value of $a_l$ in the estimation of $\lambda_{\rm KH}$ to simply $\Delta_f$. Therefore, in practice 
Eq.~\eqref{eq:wavelength:khi} is evaluated as
\begin{equation}
\lambda_{\rm KH} \approx 15.8 \Delta_f.
\end{equation}

Finally, we use  the \textit{Strang splitting} method \citep{Strang:1968} to solve Eq.~\eqref{energy_dens_ev}, since this partial differential equation is generally stiff, due to the exponential growth of the quantities. 

\subsubsection{Calibration of the coefficients of the MInIT model}
\label{sec:Calib}

\begin{figure}
    \centering
    \includegraphics[width=\linewidth]{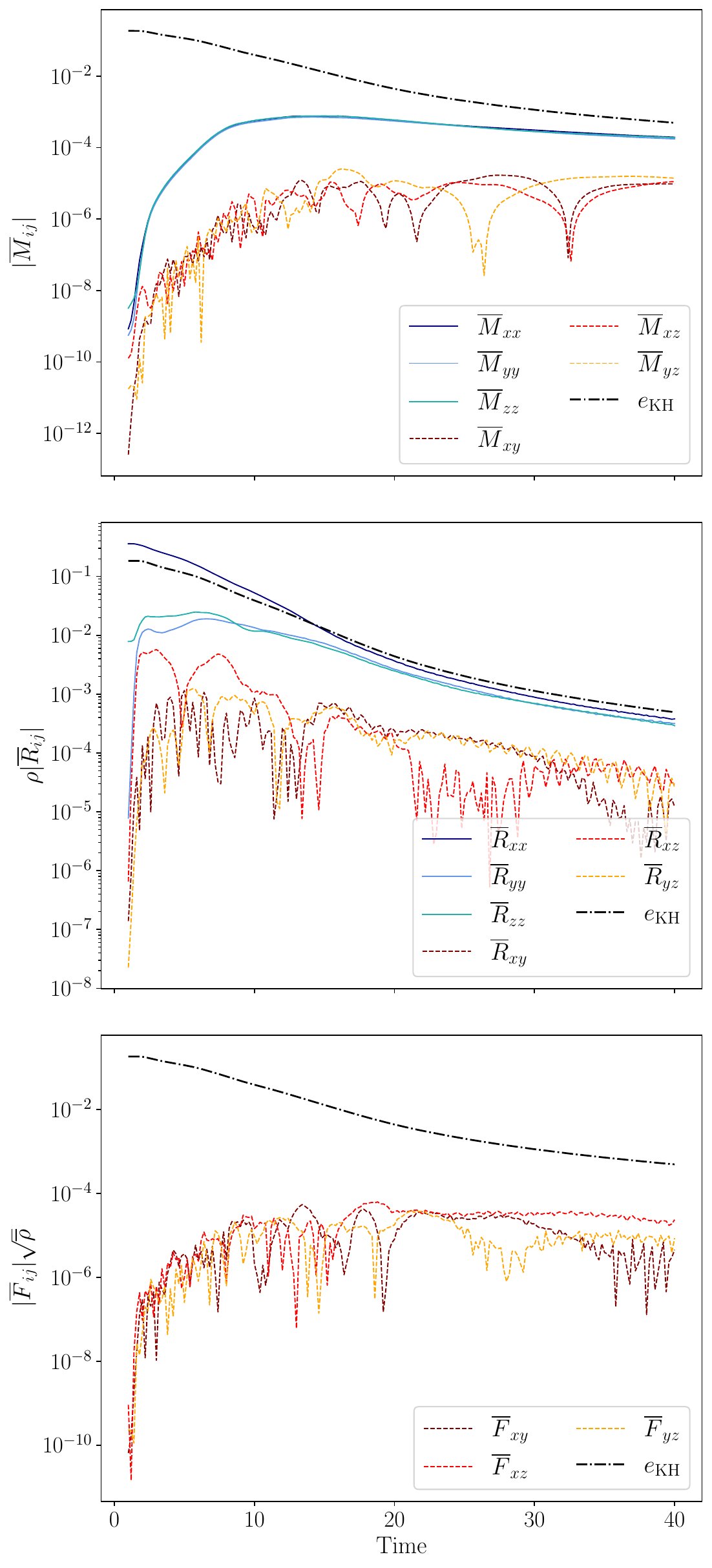}
    \caption{Time evolution of the components of the Maxwell (top), Reynolds (middle) and Faraday (bottom) stress tensors for the simulation KH-H1. The Reynolds and the Faraday stress tensors include factors with the mass density to have the same dimensions that $e_{\rm KH}$.}
    \label{fig:stresses_KH-H1}
\end{figure}

We compute the coefficients $\alpha^{\rm KH}_{ij}$, $\beta^{\rm KH}_{ij}$ and $\gamma^{\rm KH}_{ij}$ that appear in Eqs.~\eqref{closure_coeffs} in a similar way as it was done in \cite{Miravet:2022}, where the stress tensors were obtained by averaging over the whole box and over a representative time of the simulation. In the KHI case we perform averages over smaller boxes, with size similar to the wavelength of the unstable mode (Eq.~\eqref{eq:wavelength:khi}), since we assume that we are able to resolve this initial fastest growing mode. After that, we obtain a single value of the coefficients by averaging over all the boxes of size $\approx \lambda_{\rm KH}$. One of the assumptions of the MInIT model is the proportionality between the different stresses and $e_{\rm KH}$
(see Section~\ref{sec:closure}), which also implies a proportionality among the different stresses themselves. Fig.~\ref{fig:stresses_KH-H1} shows that this proportionality is actually only achieved at later times in the simulation, when turbulence is fully developed. For early times the components of the Reynolds stress that involve $v_x$ start from large values and do not have an exponential growth, due to the fact that this component represents the shear that is decreasing with time. Therefore, we will only consider late simulation times ($t>20$) for our calculations. This might be regarded as a limitation of the model. However, this drawback only affects during very short transients, since the exponential growth happens on a very short timescale compared to the duration of the simulation.

\begin{table}
    \centering
\begin{tabular}{|c c c c|}
\hline
     & $\alpha_{ij}^{\rm KH}$ & $\beta_{ij}^{\rm KH}$ & $\gamma_{ij}^{\rm KH}$   \\
    \hline 
    \hline
   $ xx $ & $0.7 \pm 0.3$ & $0.74\pm 0.05$ & -   \\
    \hline
    $yy$ & $0.64 \pm 0.18$ & $0.64\pm 0.03 $ & -   \\
    \hline
    $zz$ & $0.64 \pm 0.19$ & $0.61 \pm 0.04$ & -   \\
    \hline
    $xy$ & $-0.01 \pm 0.03$ & $-0.01 \pm 0.03$ & $-0.010 \pm 0.017$ \\
    \hline
    $xz$ & $-0.02 \pm 0.03$ & $-0.041 \pm 0.035 $ & $-0.01 \pm 0.03$ \\
    \hline
    $yz$ & $-0.026 \pm 0.025$ & $-0.02 \pm 0.03$ & $-0.008 \pm 0.015$ \\
    \hline 
\end{tabular}
    \caption{Time average of the coefficients of the closure equations of our model, Eqs.~\eqref{closure_coeffs}. The uncertainty arises from both the time average of the coefficients themselves and the spatial average of the stress tensors used to calculate the coefficients.}
    \label{tab:table_coeffs}
\end{table}
 
The estimated values of the coefficients that we obtain are reported in Table~\ref{tab:table_coeffs}. These values are the average of those corresponding to simulations KH-H1 and KH-H2. The initial magnetic field strength of the run KH-H3 is so large that the magnetic energy dominates over the kinetic energy, and the results differ considerably from the other two cases. We believe that this is not the regime we are interested in in the context of BNS mergers since magnetic fields are small ($10^{10}$ - $10^{12}$~G) when the merger occurs \citep{Lorimer2008}. Therefore, we compute the coefficients using only the simulations with lower initial magnetic field. We find that the diagonal components of $\alpha^{\rm KH}_{ij}$ and $\beta^{\rm KH}_{ij}$ are all positive. All non-diagonal components of the stresses are much smaller than the diagonal ones, and most of them are also compatible with zero.
 
\begin{figure*}
\centering
   \includegraphics[width=0.8\textwidth]{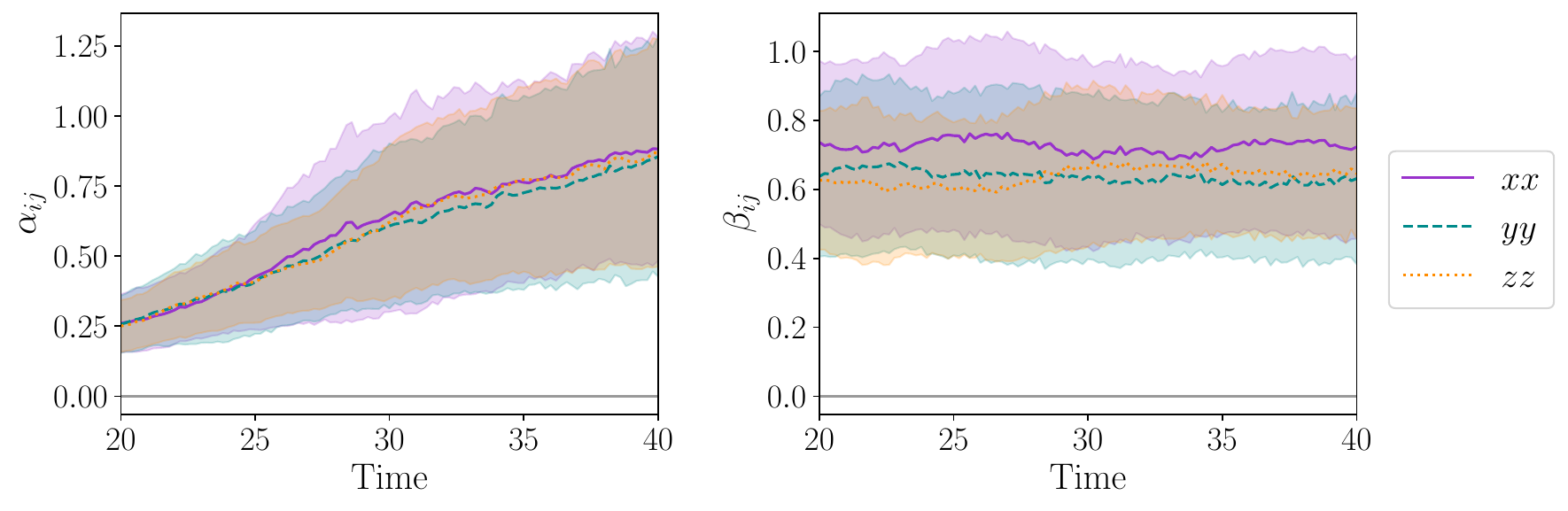}
    \caption{Time evolution of the diagonal components of the $\alpha^{\rm KH}$ and $\beta^{\rm KH}$ coefficients from simulation KH-H1. The shadows represent the standard deviation that arises from the average over the whole simulation box. Note that the values of each component are consistently time-independent when considering late times.}
   \label{fig:coeffs-diag}
\end{figure*}
 
\begin{figure*}
\centering
   \includegraphics[width=\textwidth]{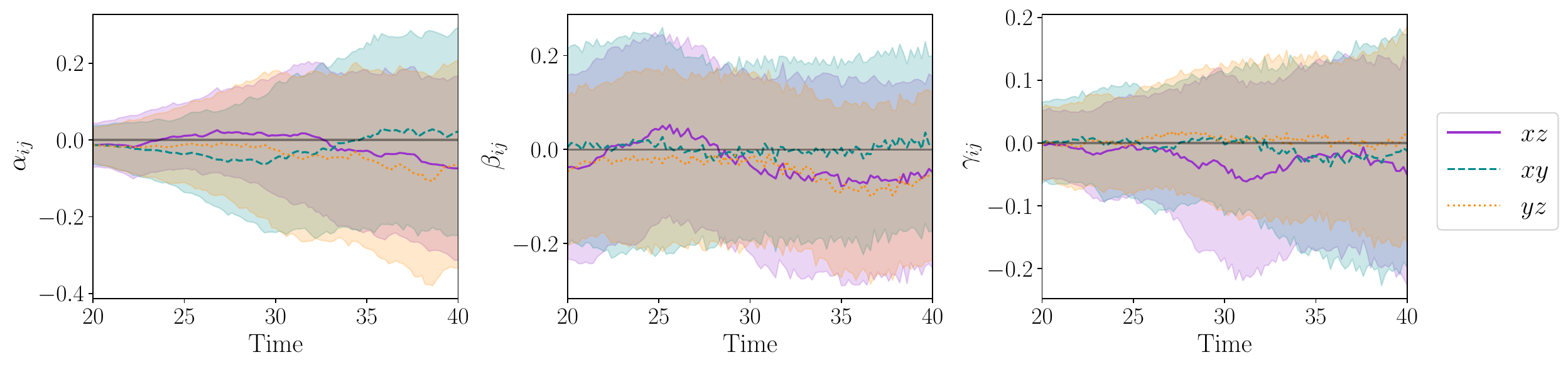}
    \caption{Time evolution of the non-diagonal components of the  $\alpha^{\rm KH}$, $\beta^{\rm KH}$ and $\gamma^{\rm KH}$ coefficients. As in Fig.~\ref{fig:coeffs-diag}, the shadows represent the standard deviation that arises from the average over the whole simulation.}
   \label{fig:coeffs-non-diag}
\end{figure*} 
 
By averaging over space and time (for $t\gtrsim 20$), we are assuming that there are no statistical spatial or temporal variations in the averaging domain and all samples (points and times) are representative of the same quantity we want to measure. We depict the temporal behaviour of the coefficients in Figs.~\ref{fig:coeffs-diag} and \ref{fig:coeffs-non-diag} for the diagonal and non-diagonal components of the coefficients, respectively, computed only from spatial averages. The shaded regions in both figures represent the standard deviation from the spatial average of the stress tensors. The coefficients almost do not change in time (except $\alpha^{\rm KH}_{ij}$), and the uncertainty comes from their spatial variability. The diagonal $\alpha^{\rm KH}_{ij}$ coefficients have larger uncertainties due to their increase in time (see Fig.~\ref{fig:coeffs-diag}).  This increase is due to the fact that the turbulence is decaying (and the turbulent kinetic energy goes down) while the turbulent magnetic energy slowly increases up to equipartition with the kinetic component. 

\subsubsection{Optimization of the $C$ and $A_{\gamma}$ parameters of the  model}
\label{sec:OPTC}

\begin{figure*}
\centering
   \includegraphics[width=\textwidth]{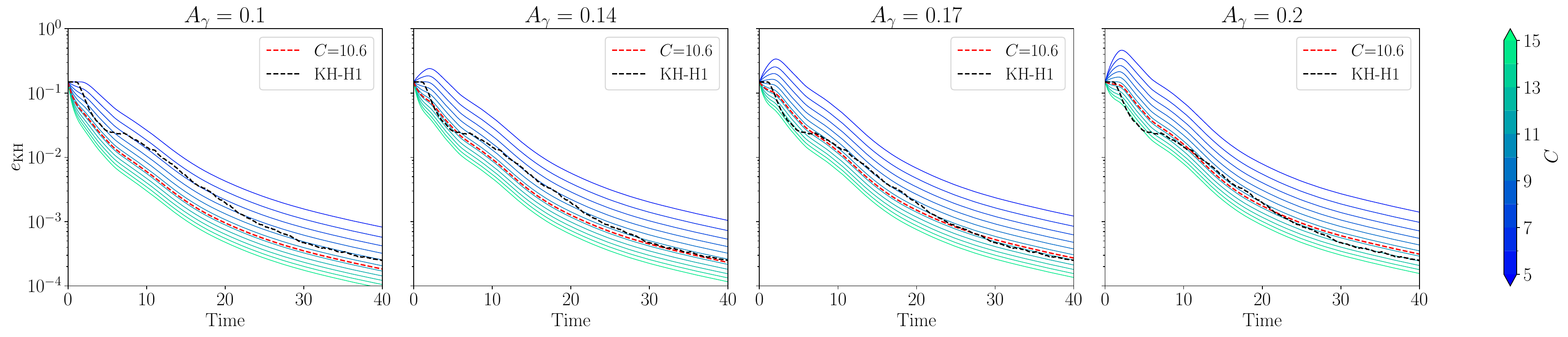}
    \caption{Time evolution of $e_{\rm KH}$ (solid lines) using the MInIT model for different values of constants $C$ and $A_{\gamma}$. The black dashed line shows the data from the simulation KH-H1 and the red dashed line highlights the case with the optimal value for $C$.}
   \label{fig:en_dens_ev}
\end{figure*}

After the values of the coefficients of the closure relations have been obtained
we still need to fix the values of two additional free parameters: the dimensionless constant $C$ from Eq.~\eqref{kolgomorov} and the factor $A_{\gamma}$ from Eq.~\eqref{grate_num}. From Eq.~\eqref{growth_rate_khi} we know already we can estimate $A_{\gamma} = 0.14$. However, before adopting this particular value, we explore the dependence of our results with other possible values for this constant. To do so we compare in Fig.~\ref{fig:en_dens_ev} the time evolution of $e_{\rm KH}$ as computed directly from the simulation KH-H1 with its value computed from the MInIT model for different values of $C$ and $A_{\gamma}$. Our goal is to find the optimal values of these two constants matching the evolution given by the simulation.  Fig.~\ref{fig:en_dens_ev} shows that for a wide range of values of $A_\gamma$ it is always possible to find a value of $C$ that yields an evolution of $e_{\rm KH}$ close to the simulation values. Therefore, given the weak dependence of the results with $A_\gamma$ we fix it to the value discussed in Section~\ref{sec:growthrates}, i.e.~$A_\gamma=0.14$ and proceed with the optimization for $C$ alone. 
 
To optimize the value of $C$, we minimize the differences between the results from the direct numerical simulation and our model. These differences are estimated using the $L_2$ (relative error) norm, 
\begin{equation}\label{l2norm}
    L_2 = \sqrt{\frac{1}{2}\sum_i(x^i_{\rm s}-x^i_{\rm m})^2\Bigg(\frac{1}{\sum_j{(x^j_{\rm s})^2}}+\frac{1}{\sum_j{(x^j_{\rm m})^2}}\Bigg)}\,,
\end{equation}
where $x^i_{\rm s}$ are the data from the numerical simulation, and $x^i_{\rm m}$ are the data from the model. As in \cite{Miravet:2022} we sum over all spatial values and components of the Maxwell and Reynolds stress tensors for each time iteration, and then we compute the root mean square over time to obtain a single value. We do not consider the Faraday stress tensor components because their values are considerably lower than those of the other stresses.
The resulting optimal value is:
\begin{equation}\label{c_optimal}
    C_{\rm opt} = 10.6 \pm 1.5 \,.
\end{equation}
The upper and lower bounds indicate $10\%$ variations of the minimized $L_2$-norm. It is worth noting that this value is fully compatible with the one we obtained for the MRI case in \cite{Miravet:2022}. Our final single value of $C_{\rm opt}$ has been obtained after averaging over different filter sizes using both the KH-H1 and KH-H2 simulations. For the same reason of Section~\ref{sec:Calib} the simulation KH-H3 was discarded in the calibration of $C$. The filter sizes used span from $\Delta_f = 40\Delta$ to $\Delta_f = 70\Delta$, an interval which is inside the inertial range of scales. 
Table~\ref{tab:table_factor_sat} reports the optimal values of $C$ for different filter sizes. The sizes that we use are also valid for the simulations with lower resolution, since they correspond to scales inside the inertial range in these cases too.

\begin{table}
  \centering
  \begin{tabular}{|c|c c c|}
\hline
    \multicolumn{4}{c}{\textbf{Optimal values of the $C$ parameter}} \\
    \hline
    \hline
       &  KH-H1 & KH-H2 &  MEAN \\
    \hline
    $S_f = 40$ & $9.3 \pm 1.5 $ &  $8.3 \pm 1.5$ & $8.8 \pm 1.5 $  \\
    \hline
   $S_f = 48$ & $10.7 \pm 1.5 $ & $9.3 \pm 1.5 $ & $10.0 \pm 1.5 $ \\
    \hline
    $S_f = 56$ & $11.7 \pm 1.5 $ & $10.3 \pm 1.5 $ & $11.0 \pm 1.5 $\\
    \hline
    
\end{tabular}
\caption{Optimal values of the parameter $C$ that minimize the $L_2$-norm of the Maxwell and Reynolds stress tensors for different filter sizes. The last column reports the mean values of the two.}
\label{tab:table_factor_sat}
\end{table}

\subsection{An a-priori test of the model}

\begin{figure*}
    \centering
    \includegraphics[width = \linewidth]{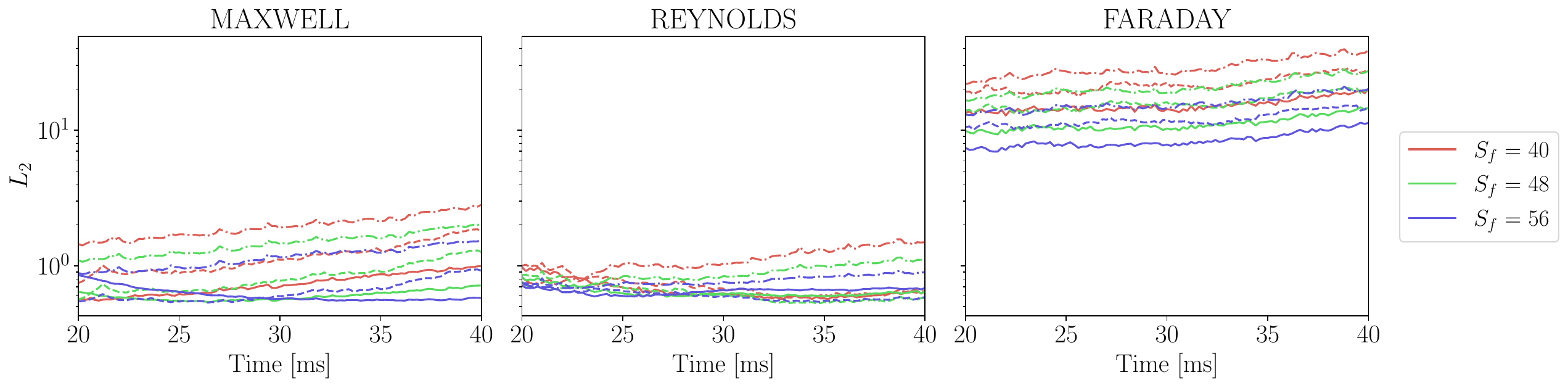}
    \caption{A-priori test of the MInIT model: time evolution of the $L_2$-norm of the Maxwell (left), Reynolds (middle), and Faraday (right) stresses. The norm represents the difference between the quantities computed from the model and the ones from the output of the simulations, as shown in Eq.~\eqref{l2norm}. We use data from simulations KH-H1 (solid lines), KH-H2 (dashed lines) and KH-H3 (dash-dotted lines). The $L_2$-norm is $\sim 1$ for the Maxwell and Reynolds stress tensors while it is $\sim 10$ for the Faraday tensor.}
    \label{fig:l2_time_ev}
\end{figure*}

To close our investigation we next present an a-priori test of our MInIT model for the KHI, paralleling what we did in~\cite{Miravet:2022} for the MRI. To do so we compute the $L_2$-norm of the difference between the numerical data, $x^i_{\rm s}$, and the data obtained with the evolution equations of the model, $x^i_{\rm m}$, using the optimized coefficients.  Figure~\ref{fig:l2_time_ev} shows the time evolution of the norm for the three stress tensors and different filter sizes, for the simulations with the highest resolutions. 
In the cases with the lowest initial magnetic fields (simulations KH-H1 and KH-H2) the $L_2$-norm of the 
Maxwell and Reynolds stresses lays below $\sim 1$ for all filters and at all times. For the highest magnetic field simulation (KH-H3) the norm is slightly above $1$ for the Maxwell stress, but still around $1$ for the Reynolds stress. This result implies that the model is able to give an order-of-magnitude estimate of the evolution of the Reynolds and Maxwell stresses. However, the norm for the Faraday tensor is higher, around 10, for the same reason it also was in the MRI case~\citep{Miravet:2022}. The discrepancy is higher for the Faraday tensor simply because the time and spatial averages of the proportionality coefficients render difficult to capture the variability and the change of sign of this tensor components. All in all, the comparison of this a-priori test with our previous results for the MRI~\citep{Miravet:2022} are consistent and promising. In our former work, the values we obtained for the $L_2$ norm of the Maxwell and Reynolds stresses were also $\sim 1$ for both the MInIT model and the gradient sub-grid model,  and around 10 for the Faraday tensor at late times.

\begin{figure*}
    \centering
    \includegraphics[width = \linewidth]{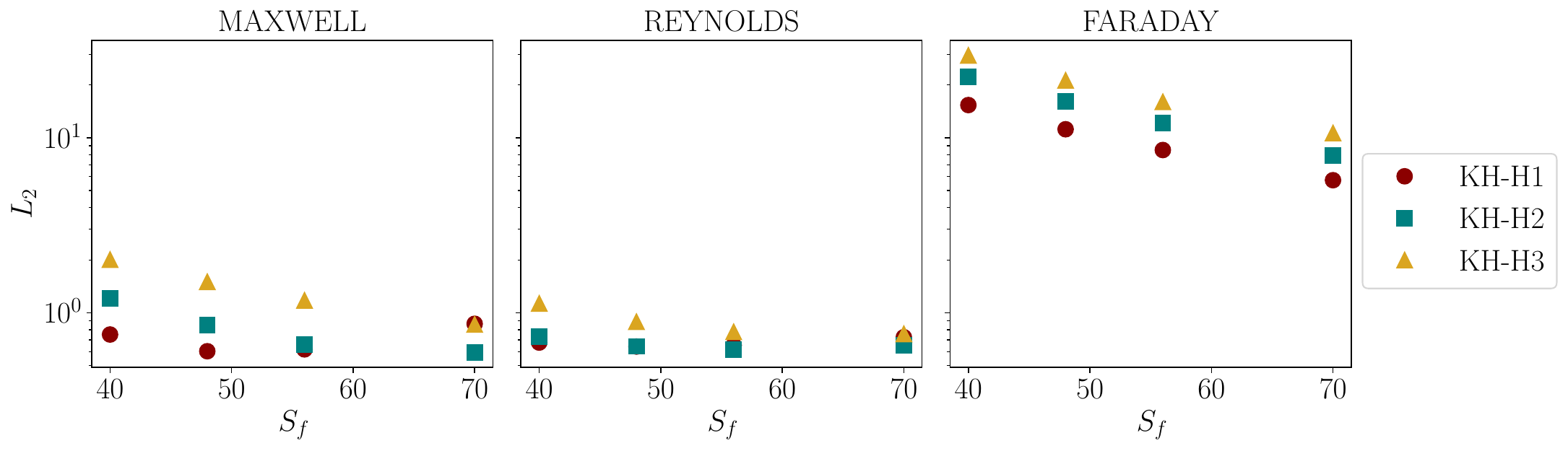}
    \caption{$L_2$-norm of the Maxwell Left), Reynolds (middle) and Faraday (right) stress tensors of the MInIT model for different filter sizes, computed over space and time-averaged. Again, the norm represents the difference between our model and the output of the simulations, as shown in Eq.~\eqref{l2norm}. Colours correspond to simulations with different initial magnetic field, as indicated in the legend.}
    \label{fig:l2_sf_MRF}
\end{figure*}

Finally, we depict in Figure~\ref{fig:l2_sf_MRF} the dependence of the model performance on the filter size, by computing the root mean square of the $L_2$-norm over time. As we found for the MRI, the norm is almost independent of the size of the filter. Moreover, it attains slightly lower values for larger filter sizes. We added here a larger filter size, $S_f = 70$, that corresponds to the inertial range of scales for the simulations with highest resolution (KH-H1, KH-H2 and KH-H3).

\section{Conclusions}
\label{sect: sec5}

There are many examples of astrophysical systems for which the proper numerical modelling of their dynamics is severely hampered by insufficient computational resolution. A particularly good example is that of fluids at high Reynolds number where the available resolution of grid-based codes for direct numerical simulations is usually inadequate to capture the physics at all scales. Relativistic astrophysical systems such as binary neutron star mergers or core collapse supernovae may therefore be affected by a deficient modelling when attempting to numerically resolve the amplification of magnetic fields or the transition to fully fledged turbulence from the growth of dynamical instabilities at small scales. However, being able to resolve the Kelvin-Helmholtz instability and the magnetorotational instability is essential for a faithful representation of the post-merger evolution of the remnant in binary neutron star mergers.  

Notwithstanding the significant progress achieved through direct numerical simulations, it is still not possible to capture all the (extreme) physics characteristic of those astrophysical systems. An alternative to performing costly high-resolution simulations are sub-grid models that attempt to express the effects of the turbulent scales in terms of resolved quantities. Several of those models have been recently implemented in numerical simulations of BNS mergers, such as the $\alpha$-viscosity and the gradient models~\citep{producing_magnetar_fields_giacomazzo,general_rel_kiuchi_shibata,Shibata:2021,extension_subgrid_vigano,paper_gradient_test,Vigano:2020,Palenzuela:2022,Aguilera-Miret:2022}.

In this paper we have presented an extension of our new sub-grid model MInIT, an MHD-instability-induced-turbulence mean-field model we first proposed in \cite{Miravet:2022}. In our previous investigation we assessed MInIT in the context of the modelling of turbulence generated by the magnetorotational instability. In the present work we have focused in evaluating MInIT for resolving KHI-induced turbulence. The main appeal of our model is that it is physically motivated, being based on the estimation of the turbulent stress tensors through the modelling of the temporal evolution of a turbulent energy density of the instability. The model consists in evolution equations for the turbulent energy densities plus a closure relation that allows to compute all turbulent stresses as proportional to those energy densities. Moreover, it takes into account the fastest growing mode of the instability and the dissipation of the energy at the end of the Kolmogorov cascade. While in our previous work on the MRI MInIT required two evolution equations, one for the MRI and a second one for the turbulence itself, for the modelling of the KHI only one evolution equation has been needed.   

The MInIT model depends on several coefficients and constants that must be calibrated. This has been done using numerical box simulations of the KHI in which part of the turbulent cascade is resolved and convergence is achieved, at least in the main global features. We have observed that the calibration of the constants does not depend strongly on the numerical resolution used in the simulations or on the initial value of the seed magnetic field, as long as it is weak. For high magnetic fields, however, some deviations have been found. 

We have assessed the performance of the MInIT model for the KHI in the same way we previously did for the MRI in \cite{Miravet:2022}, i.e.~via an a-priori test. We have used data from direct numerical simulations and applied a box filter to see the difference between the quantities given by the model and the filtered ones. To do so, the $L_2$ relative error norm was used to quantitatively compare with the numerical data. We have found that the values of the $L_2$ norm  lay below $\sim 1$ for the Maxwell and Reynolds stresses. Therefore, we are able to obtain order-of-magnitude accurate estimates of these stresses with our model. In comparison with the MRI version, the MInIT model for the KHI performs slightly better, since the data from the simulation and the modeled stresses have been found to differ less. More precisely, the discrepancy is less than an order of magnitude, which is an achievement worth mentioning given the simplicity of the model. 

For the Faraday stress tensor the results are worse; the typical values of the $L_2$ relative error are around $10$ at most. 
This could be explained by the fact that the model coefficients we found for the Faraday tensor components are actually compatible with zero (a similar result was obtained in \cite{Miravet:2022} for the MRI). This means that our model is compatible with setting the Faraday stress to zero. However, in reality the Faraday tensor components should probably have some small (but non-zero) value. Its magnitude is actually relevant since it impacts the ability of the model to develop large-scale turbulent dynamos that may operate in BNS mergers and magnetar formation. 

In order to obtain more accurate measurements of the Faraday stress one should build more statistics. Increasing the computational time of the simulations might help to some extent. However, since our simulations are effectively ones of decaying turbulence, their time extension is limited by nature. Alternatively, one could explore the case of driven turbulence which shares some similarities with the box simulations displayed in this work. In this case, the sustained nature of the turbulence would allow for longer simulations and build more statistics.

Furthermore, we have also found that our sub-grid model shows no strong dependence on the filter size $S_f$. This is something to keep in mind, since an ideal sub-grid model should work properly in the limit $S_f \rightarrow \infty$. Simulations with different initial magnetic field amplitudes (always in the regime of weak amplitude) yield similar values of the $L_2$-norm, which is reassuring.

Together with the version of MInIT developed for the MRI in~\cite{Miravet:2022}, 
the sub-grid model presented in this work could eventually be applied to different astrophysical systems, in particular in the study of the dynamics of the merger and post-merger phases in BNS coalescences. The early phase of the merger, when both neutron stars make contact, is characterized by the excitation of the KHI, which can lead to a substantial amplification of the magnetic field of the system, provided turbulence is properly captured. This is something we plan to investigate with the model put forward in this paper. However, before doing that some extra work should be carried out in the present version of the model. In particular, we plan to relax some of the implicit assumptions of the model, such as considering weak magnetic field seeds or large Reynolds numbers. Our future goals are the generalization of MInIT to account for a wider range of physical conditions and its implementation in numerical simulations using a general-relativistic framework. The generalisation to curved spacetimes will require the inclusion of metric-dependent terms in the right-hand-side of equations for the
unresolved turbulent energy densities as well as for the large-scale fluid and magnetic field.  The additional terms do not change the basic character of the hyperbolic system and can be treated using standard techniques for dealing with local source terms in general relativistic MHD. The details of such an implementation will be addressed in future work.

\section*{Acknowledgements}

The authors thank Milton Ruiz for useful discussions. Work supported by the Spanish Agencia Estatal de Investigación (Grants PID2021-125485NB-C21 and PID2021-127495NB-I00 "SUPERNOVAE", funded by MCIN/AEI/10.13039/501100011033 and ERDF A way
of making Europe), by the Generalitat Valenciana (Prometeo grants CIPROM/2022/13 and CIPROM/2022/49), by the Astrophysics and High Energy Physics programme of the Generalitat Valenciana ASFAE/2022/03 and ASFAE/2022/026 funded by MCIN and the European Union NextGenerationEU (PRTR-C17.I1). Further support is provided by the EU's Horizon 2020 research and innovation (RISE) programme H2020-MSCA-RISE-2017 (FunFiCO-777740), and  by  the  European Horizon  Europe  staff  exchange  (SE)  programme HORIZON-MSCA-2021-SE-01 (NewFunFiCO-101086251). MMT acknowledges support by the Ministerio de Universidades de España (Spanish Ministry of Universities) through the ``Ayuda para la Formación de Profesorado Universitario" (FPU) No. FPU19/01750. MO acknowledges support from the Spanish Ministry of Science, Innovation and Universities via the Ramón y Cajal programme (RYC2018-024938-I). This work has used the following open-source packages: \textsc{NumPy} \citep{harris:2020}, \textsc{SciPy} \citep{scipy:2020} and \textsc{Matplotlib} \citep{Hunter:2007}.

\section*{Data Availability}

The data underlying this article will be shared on reasonable request to the corresponding author.


\bibliographystyle{mnras}
\bibliography{KHI} 





\bsp	
\label{lastpage}
\end{document}